# Complexity science and intentional systems

*Educational Research Review* (forthcoming)

Loet Leydesdorff
Amsterdam School of Communications Research (ASCoR),
University of Amsterdam, Kloveniersburgwal 48, 1012 CX Amsterdam;
loet@leydesdorff.net ; http://www.leydesdorff.net

In their position paper entitled "Towards a new, complexity science of learning and education," Jörg *et al.* (2007) argue that educational research is in crisis. In their opinion, the transdisciplinary and interdiscursive approach of complexity science with its orientation towards self-organization, emergence, and potentiality provides new modes of inquiry, a new lexicon and assessment practices that can be used to overcome the current crisis. In this contribution, I elaborate on how complexity science can further be developed for understanding the dynamics of intentions and the communication of meaning as these are central to the social-scientific enterprise.

Under the denominator of "complexity science" a number of physicists, biologists, and mathematicians have proposed "self-organization" as a metaphor. "Self-organization," however, has a meaning in the context of Prigogine's (1980) thermodynamics of far-from-equilibrium systems that differs from its use in Maturana & Varela's (1984) neurophysiology-based model of *autopoiesis*. Luhmann (1986) proposed using the latter model to analyse the communication of meaning in social and psychological systems. The distinction between social and psychological systems was based on Husserl's (1929) philosophy, but radicalized by Luhmann to the extent that these two types of systems are considered as operationally closed and therefore as constituting environments for each other. In other words, social systems can be expected to process meaning differently from psychological systems.

The cybernetic model of self-organization may have its origins in biology or physics, but the crucial question is whether the metaphor helps to explain problems and puzzles in the system(s) under study (Holland, 1998). Unlike biology, the social sciences study intentional subjects and their social configurations. The non-linear dynamics of meaning are hitherto poorly understood as a subject of complexity science. Meaning is provided from the perspective of hindsight, and thus the arrow of time is locally reversed (Coveney & Highfield, 1990; Leydesdorff, 1994; Urry, 2003; Mackenzie, 2001). This may reduce the uncertainty that would otherwise be expected to increase because the Second Law is valid both for thermodynamics and for the dynamics of probabilistic entropy (Theil, 1972).

The mechanism of providing meaning can be modeled using the theory of anticipatory systems (Rosen, 1985). An anticipatory system is a system that is able to entertain one or more models of itself. The model provides the modeled system with specific meaning. Dubois (1998) found a way to formalize this as an *incursive* equation. Using these equations, a distinction can be made between weakly and strongly anticipatory systems.

The latter are able not only to model themselves, but also to co-construct their next future states.

In this context, I proposed using this distinction to model the difference between psychological and social systems: while psychological systems are able to entertain models of themselves, social systems are able to co-construct their own next states, for example, in the case of techno-economic co-evolutions (Leydesdorff, 2008). Using Dubois's equations, it is possible to derive formulations for the three levels at which meaning can be communicated according to Luhmann (1997): interaction, organization, and self-organization. However, it follows from these equations that the system would accumulate complexity if agency did not step in to make selective choices. The social system can therefore be considered as semi-autopoietic: the further development of the system remains dependent on agency to co-evolve, for example, in terms of communicative competencies (Habermas, 1981).

Within Luhmann's theory, this additional coupling between agents and structures can be appreciated as "interpenetration" (Parsons, 1968; Luhmann, 2002). Unlike the biological mechanism of structural coupling and operational closure, social and psychological systems have access to each other's operations. This additional degree of freedom can be considered as grounded in the emergence of human language as an evolutionary step (Leydesdorff, 2000). The controversy signaled by Habermas (1987, at p. 385) between "linguistically generated intersubjectivity" and "self-referentially closed systems" can thus be considered as a puzzle which complex systems theory may be able to solve.
return